\documentclass[10pt, conference]{IEEEtran}
\IEEEoverridecommandlockouts

\usepackage[T1]{fontenc}
\usepackage{graphicx}
\usepackage{xcolor}
\usepackage{amsmath, amssymb, amsfonts}
\usepackage{textcomp}
\usepackage{hyperref}
\usepackage{url}
\usepackage{gensymb}

\usepackage{booktabs}
\usepackage{multirow}
\usepackage{caption}
\usepackage{subfigure}

\usepackage{balance}
\usepackage{float}
\usepackage{adjustbox}
\usepackage{makecell}
\usepackage{colortbl}
\usepackage{siunitx}
\usepackage{orcidlink}  
\makeatletter
\renewcommand{\@thesubfigure}{\normalsize(\textbf{\alph{subfigure}})}
\makeatother

\begin{document}

\title{S-ALSA: Co-Design of Adiabatic Logic-based Sensing and Balanced Bit-Cells for Secure and Energy-Efficient MRAM}

\author{
\IEEEauthorblockN{Wu Yang\textsuperscript{1}\orcidlink{0000-0001-8857-4473}, Amit Degada\textsuperscript{2}\orcidlink{0000-0002-6335-694X}, Himanshu Thapliyal\textsuperscript{1}\orcidlink{0000-0001-9157-4517}}
\IEEEauthorblockA{\textsuperscript{1}Department of Electrical and Computer Engineering, Southern Methodist University, Dallas, TX 75205, USA}
\IEEEauthorblockA{\textsuperscript{2}Department of Electrical Engineering and Computer Science, University of Tennessee, Knoxville, TN 37996, USA}
}

\maketitle

\begin{abstract}
Magnetoresistive Random Access Memory (MRAM) technologies such as Spin-Transfer Torque (STT-MRAM) and Spin-Orbit Torque assisted (SOT-STT-MRAM) offer non-volatility and low leakage, making them attractive for IoT systems. However, conventional MRAM read circuits face two fundamental challenges: high dynamic energy consumption and vulnerability to side-channel attacks caused by data-dependent current variations in Magnetic Tunnel Junctions (MTJs). This paper presents a Secured Adiabatic Logic Sense Amplifier (S-ALSA) that addresses both challenges simultaneously through circuit-device co-design. S-ALSA combines structural current balancing via a 4T-2MTJ bit cell, which eliminates read current asymmetry at the storage level, with dynamic power equalization via adiabatic charge recovery in the sensing circuit. The proposed architecture supports both STT-MRAM and SOT-STT-MRAM. Case studies using $4$ $\times$ $4$ MRAM macros shows up to 80\% energy savings over conventional Pre-Charge Sense Amplifiers (PCSA) across IoT frequencies. Correlation Power Analysis (CPA) attacks on PRESENT-80 encryption confirm complete suppression of key leakage when S-ALSA is combined with a balanced bit cell. This work establishes a unified framework where energy efficiency and hardware security are achieved simultaneously, enabling secure and low-power IoT memory design.
\end{abstract}


\section{Introduction}\label{sec:intro}

IoT systems demand memory architectures that are both energy-efficient and secure. MRAM technologies, particularly STT-MRAM and SOT-STT-MRAM, offer non-volatility, high endurance, and CMOS compatibility~\cite{IoTcloud}, making them strong candidates for next-generation IoT platforms. STT-MRAM and SOT-STT-MRAM are now commercially available as both embedded memory in IoT SoCs and standalone board-level components, enabling instant-on, ultra-low-power operation. However, two major challenges remain unresolved:

\begin{itemize}
  \item High dynamic energy consumption during read operations, which dominate nearly 80\% of the MRAM read-write cycle.
  \item Side-channel leakage due to data-dependent MTJ current signatures, which make MRAM read circuits vulnerable to Differential Power Analysis (DPA) and Correlation Power Analysis (CPA)~\cite{STTsca, STTsca2}.
\end{itemize}

An attacker with access to the power supply pin and knowledge of the cryptographic round schedule can align power traces to specific MRAM read cycles and statistically correlate current variation to stored data, recovering secret keys. Existing countermeasures enforce balanced switching activity to reduce exploitable current difference, but at the cost of increased area and higher power~\cite{IoTsec, IoTsecurity1}, making them unsuitable for power-limited IoT devices.

In our earlier work, we proposed an adiabatic logic-based sense amplifier that reduced dynamic energy in STT-MRAM~\cite{wu2022adiabatic}. While that work demonstrated substantial energy savings, it did not address security. This paper advances that design by introducing a co-designed secure and energy-efficient MRAM sensing architecture. Adiabatic logic offers a dual benefit: (i) uniform power consumption for logic `1' and `0' increases the statistical difficulty of power profiling, and (ii) charge recycling reduces dynamic energy dissipation. IoT applications, including medical implants, wearables, and wireless sensor nodes, typically operate at low frequencies~\cite{9674039}, where adiabatic circuits are most effective since energy savings scale with clock period $T$ relative to time constant $RC$~\cite{teichmann2011adiabatic}.

The key contributions of this work are:

\begin{itemize}
  \item A Secured Adiabatic Logic Sense Amplifier (S-ALSA) that minimizes dynamic power variation through adiabatic charge recovery, supporting both STT-MRAM and SOT-STT-MRAM with 1T-1MTJ and 4T-2MTJ bit cells.
  \item A co-design framework pairing S-ALSA with a 4T-2MTJ balanced bit cell that eliminates structural current asymmetry at the storage level, providing two independent layers of side-channel protection.
  \item The first evaluation of an adiabatic MRAM sense amplifier against side-channel attacks, including current trace profiling, NED/NSD metrics, and CPA attacks on PRESENT-80~\cite{P80}.
  \item Energy and security evaluation on $4$ $\times$ $4$ MRAM macros across frequencies (2--25\,MHz), temperatures ($-25^\circ$C to $75^\circ$C), and TMR variations (150\%--250\%), validating up to 80\% energy savings.
\end{itemize}

\section{Background}\label{sec:background}

\subsection{Magnetic Tunnel Junction and MRAM}

The MTJ is a three-layer spintronic device (CoFeB/MgO/CoFeB) exhibiting two resistive states: a low-resistance Parallel (P) state ($R_P$) and a high-resistance Anti-Parallel (AP) state ($R_{AP}$), quantified by the TMR ratio (Eq.~\ref{equ:TMR_ratio}). We use Perpendicular Magnetic Anisotropy (PMA) MTJs~\cite{zhang2015compact} with $R_P = 5\,\text{k}\Omega$, $R_{AP} = 15\,\text{k}\Omega$, and TMR~$= 200\%$ unless noted.

\begin{equation}
TMR = \frac{R_{AP} - R_{P}}{R_{P}}
\label{equ:TMR_ratio}
\end{equation}

\begin{figure}[htbp]
\centering
\includegraphics[width=0.62\columnwidth]{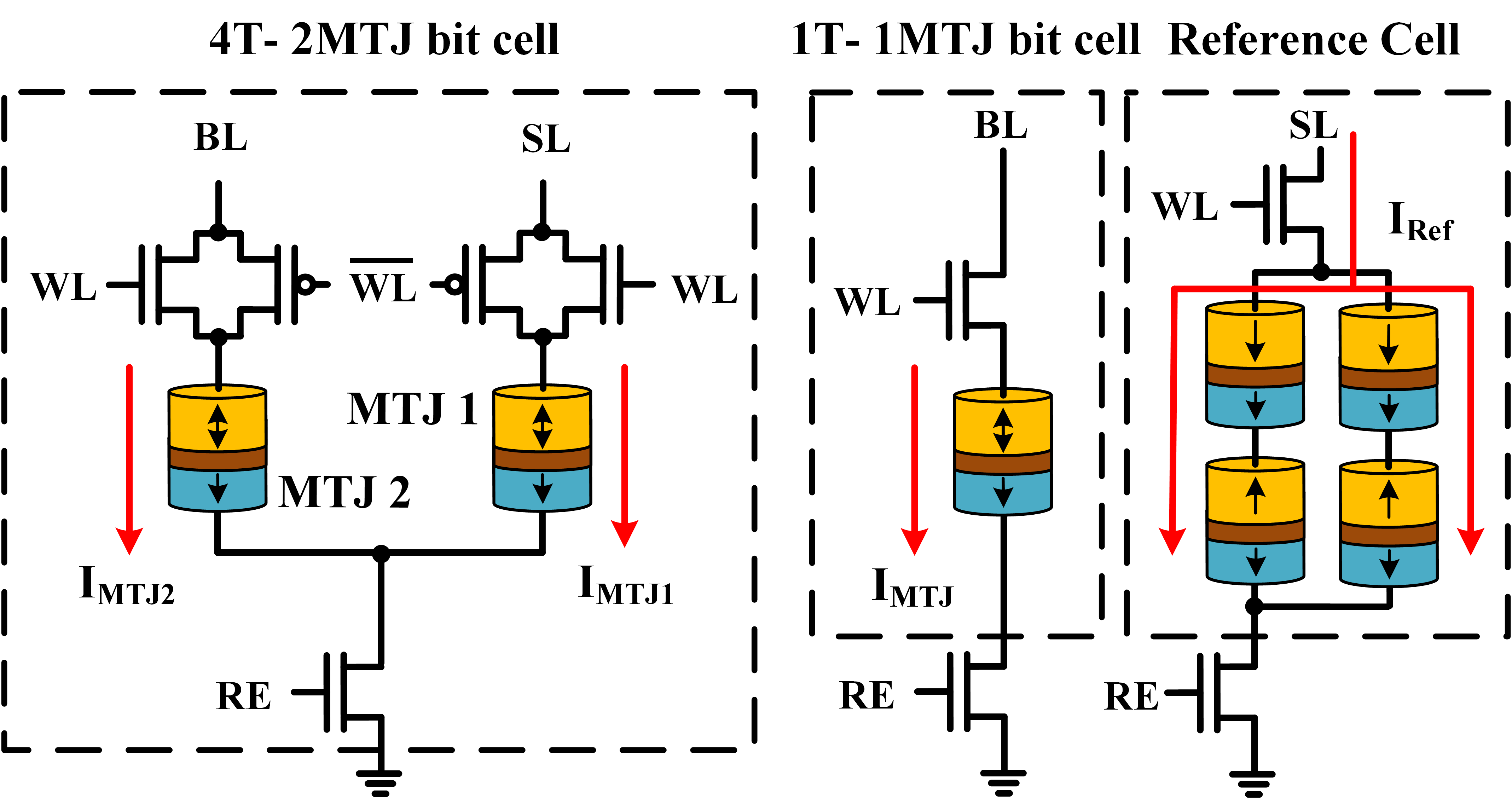}
\caption{Bit cell structures: (left) 4T-2MTJ, (middle) 1T-1MTJ, (right) Reference cell.}
\label{fig:bitcell}
\end{figure}

STT-MRAM applies spin-polarized current through the MTJ to switch its state, offering non-volatility with DRAM-like speed~\cite{li2023experimental}. SOT-STT-MRAM further uses a heavy-metal layer to inject spin-polarized current via the spin-Hall effect, enabling faster switching and better energy efficiency than STT alone~\cite{wang2018evaluation}. Both technologies support 1T-1MTJ and 4T-2MTJ bit cell configurations (Fig.~\ref{fig:bitcell}). The 1T-1MTJ cell uses a reference cell for comparison during reads, which introduces asymmetry between logic `0' and `1' read currents. This asymmetry is a key security vulnerability. The 4T-2MTJ cell stores data and its complement in two MTJs, so the sense amplifier always compares one P-state and one AP-state MTJ, structurally balancing the read current. The Pre-Charge Sense Amplifier (PCSA) is the conventional reading circuit and serves as our baseline.

\subsection{Adiabatic Logic}

Adiabatic logic uses a slowly rising/falling power clock as a constant current source for the capacitive load, enabling charge recovery rather than dissipation. Energy dissipation is:

\begin{equation}
E_{\text{adiabatic}} = \frac{RC}{T}CV_{dd}^2 \qquad E_{\text{CMOS}} = \frac{1}{2}CV_{dd}^2
\label{equ:Energy}
\end{equation}

where $C$ is load capacitance, $R$ is transistor parasitic resistance, $T$ is the clock transition period, and $V_{dd}$ is the full swing. When $T > 2RC$, adiabatic circuits dissipate less energy than conventional CMOS. IoT devices operate at low frequencies with high $T/RC$ ratios, maximizing the energy recovery benefit~\cite{9674039, teichmann2011adiabatic}.

\subsection{Power Analysis Attacks on MRAM}

\begin{figure}[htbp]
\centering
\includegraphics[width=0.62\columnwidth]{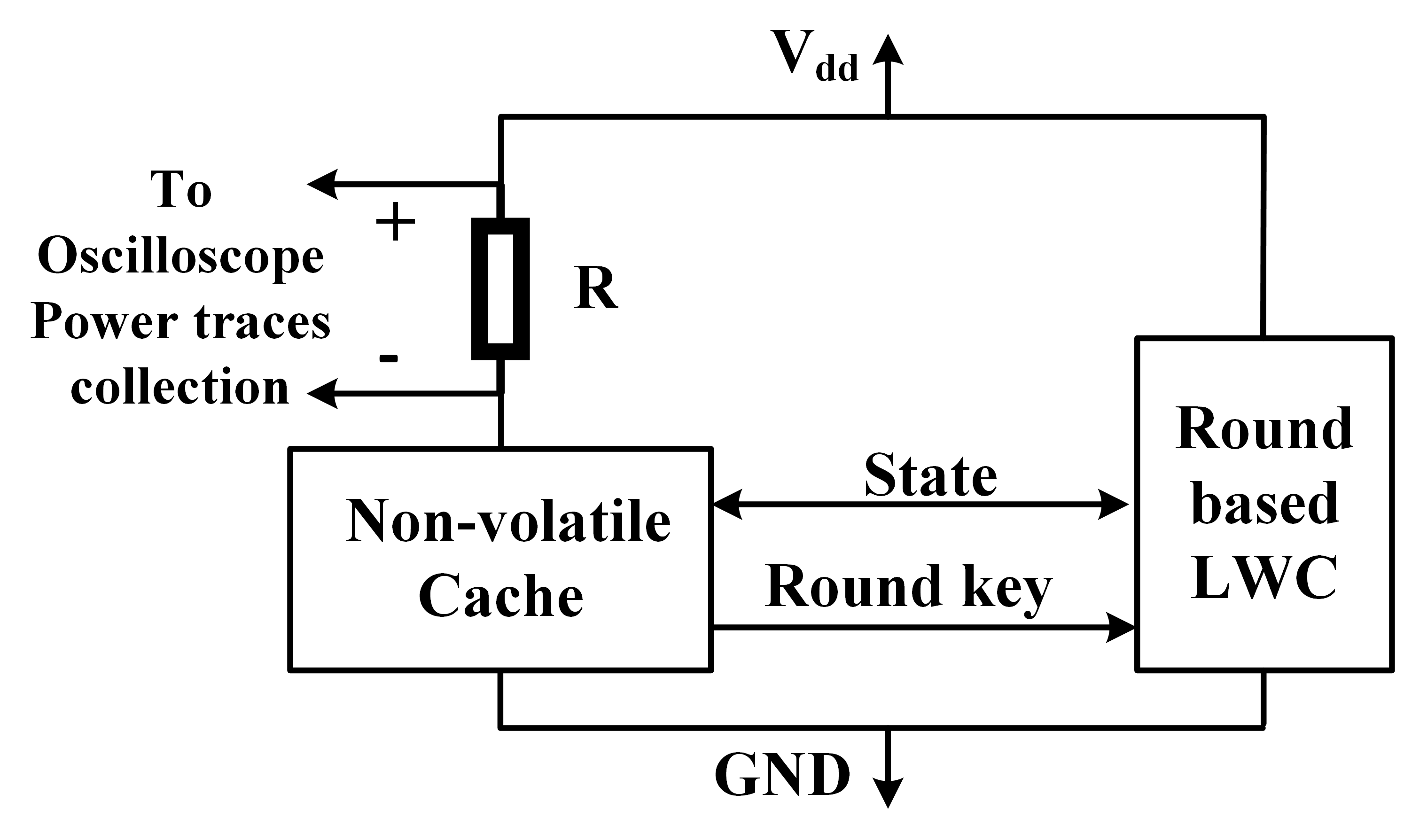}
\caption{System-level view of a power analysis attack on a round-based lightweight cipher.}
\label{fig:attack_setup}
\end{figure}

IoT devices commonly implement lightweight cryptographic ciphers (e.g., PRESENT-80~\cite{P80}) using a round-based architecture: the cipher core is reused each round, with intermediate states and round keys stored in MRAM between rounds. Fig.~\ref{fig:attack_setup} illustrates this threat model. An attacker inserts a small sense resistor ($R_s$) in series with the MRAM supply to capture current traces non-invasively. Because the MTJ exhibits asymmetric resistance, each read cycle leaks the stored data bit through the power side-channel~\cite{STTsca, STTsca2}.

DPA partitions power traces by a hypothesized intermediate bit and recovers the key through difference-of-means analysis. CPA builds a Hamming-weight power model and identifies the key guess with the highest Pearson correlation to measured traces~\cite{brier2004correlation}, requiring fewer traces than DPA. The fundamental countermeasure is to eliminate data-dependent power variation: adiabatic logic achieves this by producing an identical number of charge/discharge transitions for logic `1' and `0', directly defeating the statistical assumption both attacks rely on.

\section{Proposed S-ALSA Architecture}\label{sec:proposed}

The proposed S-ALSA is a current-mode sense amplifier consisting of: (i) back-to-back CMOS inverters (MP1-MN1, MP2-MN2) for sensing and latching, and (ii) two NMOS access transistors (MN5, MN6) connecting output nodes to the bit cell. Fig.~\ref{fig:ALSA} shows the S-ALSA interfaced with the bit cell and write circuit. A 2-phase sinusoidal power clock ($PC$) is used for simplicity. The read cycle proceeds through four sub-phases:

\begin{figure}[htbp]
\centering
\includegraphics[width=0.55\columnwidth]{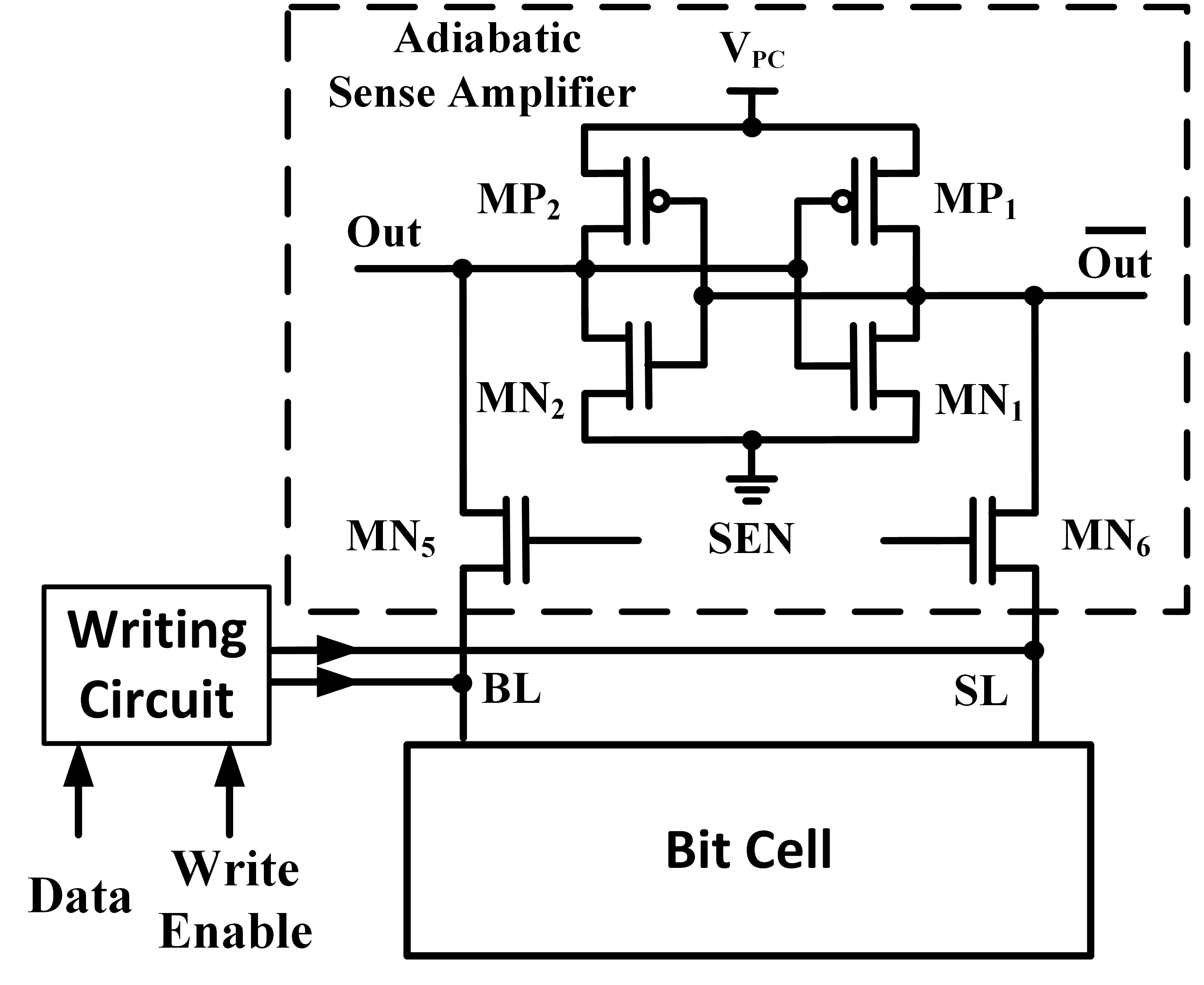}
\caption{Proposed Secured Adiabatic-Logic-Based Sense Amplifier (S-ALSA)~\cite{wu2022adiabatic}.}
\label{fig:ALSA}
\end{figure}

\textbf{(1) Preparation:} $WL$ selects the bit cell; a \textit{SENSE} pulse discharges residual charge from $Out$, $\overline{Out}$, BL, and SL to GND for balanced initial conditions.

\textbf{(2) Sensing:} As $V_{pc}$ rises, MP1 and MP2 enter the triode region. The MTJ resistance difference (AP vs.\ P state) induces a current imbalance, building a voltage differential between $Out$ and $\overline{Out}$.

\textbf{(3) Amplifying:} Once sufficient differential voltage is established, \textit{SENSE} is de-asserted, disconnecting S-ALSA from the bit cell. The cross-coupled inverters amplify the difference.

\textbf{(4) Latching/Recovery:} When $Out$ reaches $V_{thn}$, MN1 pulls $\overline{Out}$ to GND, completing the latch. As $V_{pc}$ falls, stored charge on $Out$ is recovered back to the power clock, reducing net energy dissipation.

Full waveform and switching diagrams are provided in our prior work~\cite{wu2022adiabatic}. The present work extends that design in two significant directions. First, S-ALSA is extended to support both STT-MTJ and SOT-STT-MTJ technologies across 1T-1MTJ and 4T-2MTJ bit cell configurations, enabling a direct comparison of the two most relevant MRAM technologies for IoT. Second, and more importantly, this work subjects an adiabatic MRAM reading circuit to rigorous side-channel security evaluation, establishing whether adiabatic logic's inherent power uniformity translates into measurable CPA resistance in practice.

\section{Case-Study: $4$ $\times$ $4$ MRAM Analysis}\label{sec:energy}

\subsection{The $4$ $\times$ $4$ MRAM Macro}

\begin{figure}[htbp]
\centering
\includegraphics[width=0.60\columnwidth]{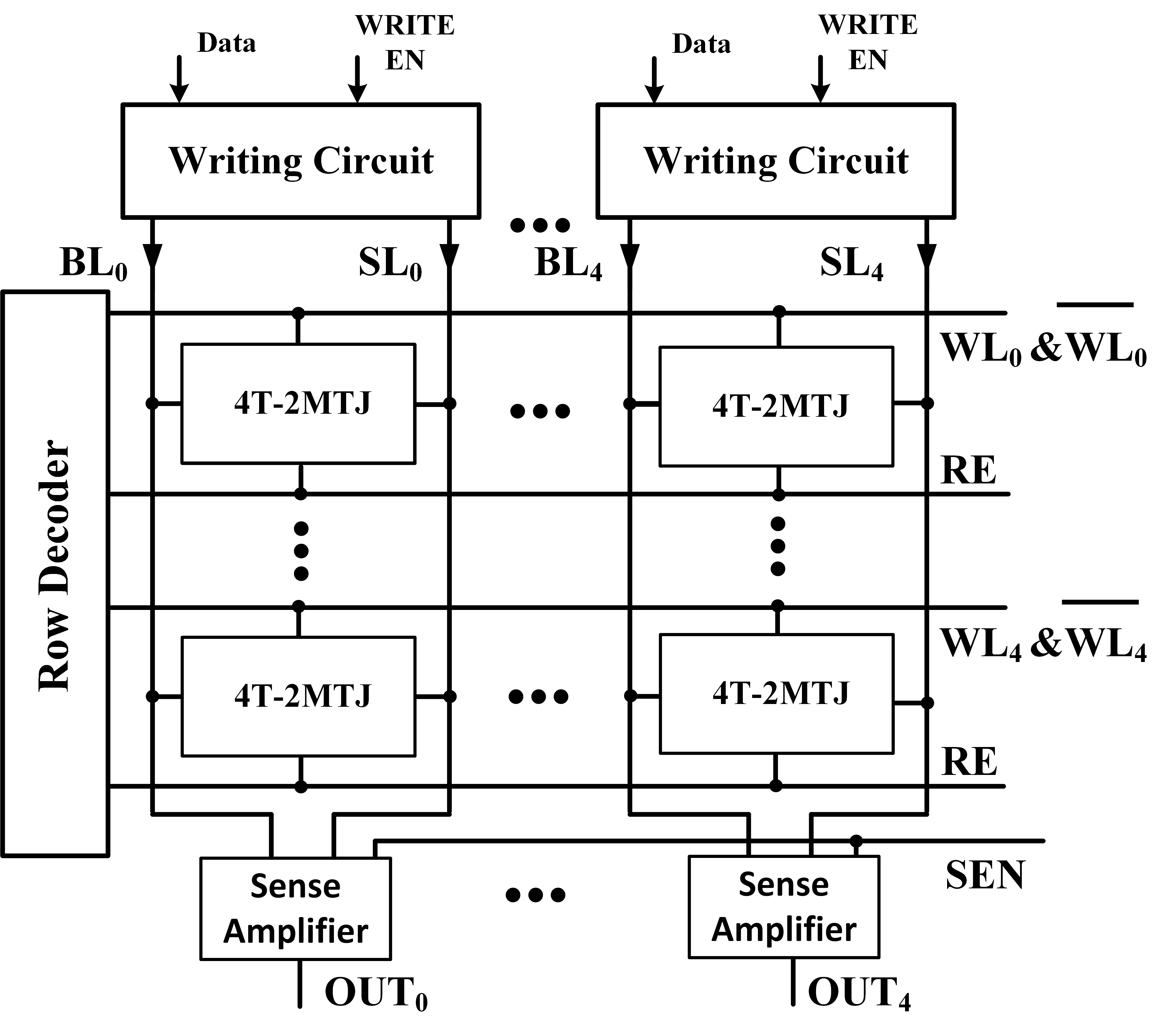}
\caption{$4$ $\times$ $4$ MRAM macro architecture (shown with 4T-2MTJ bit cells).}
\label{fig:4x4}
\end{figure}

A $4$ $\times$ $4$ MRAM macro organizes bit cells into 4 rows and 4 columns, with one sense amplifier per column, a shared write circuit per column, and a row decoder selecting a wordline for read or write (Fig.~\ref{fig:4x4}). During a read, all four sense amplifiers operate in parallel, producing a 4-bit output per cycle. The $4$ $\times$ $4$ macro is the standard MRAM evaluation vehicle: it is the smallest array exercising all peripheral circuits simultaneously, capturing realistic loading and parasitics absent in single-bit studies, and is directly scalable by tiling. Its 4-bit output also matches the nibble-wide datapath of PRESENT-80, enabling a direct CPA attack mapping over all 16 stored-value combinations. SPICE simulations used TSMC 65\,nm PDK and Verilog-A MTJ models~\cite{zhang2015compact, wang2015perpendicular} with 10\,fF load capacitance, 100 read-after-write cycles, and equal distributions of logic `1' and `0'. Both 1T-1MTJ and 4T-2MTJ bit cells were evaluated. The 1T-1MTJ cell offers higher density, while the 4T-2MTJ cell incurs approximately $3.5\times$ area overhead (4 transistors + 2 MTJs vs.\ 1 transistor + 1 MTJ) but provides balanced read currents and improved security, as analyzed in Section~\ref{sec:security}. For security-critical IoT applications, this area cost is justified by the complete elimination of data-dependent current asymmetry.

\subsection{Energy vs. Frequency}

\begin{figure}[htbp]
\centering
\subfigure[$4$ $\times$ $4$ STT-MRAM]{\includegraphics[width=0.41\columnwidth]{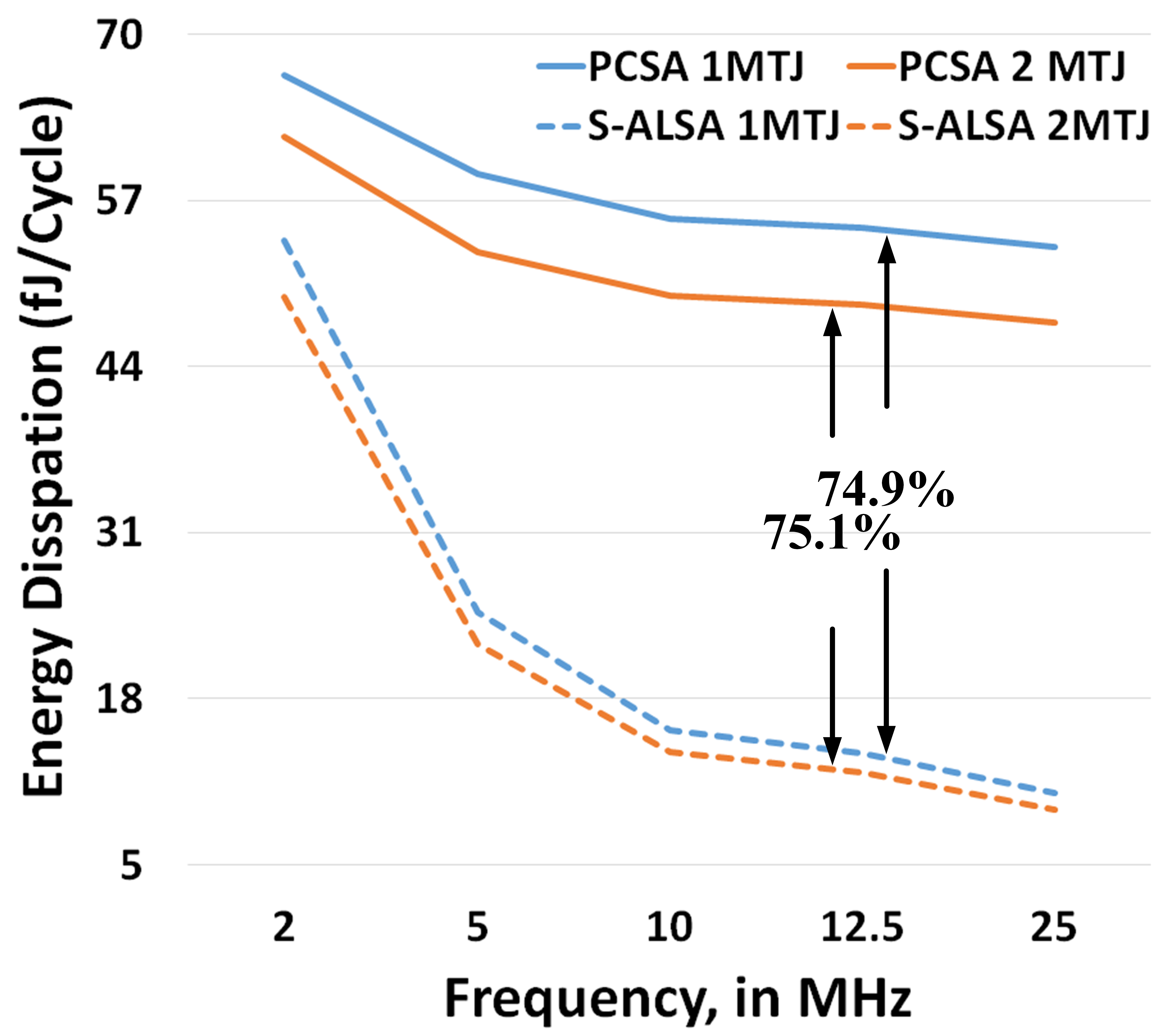}}
\subfigure[$4$ $\times$ $4$ SOT-STT-MRAM]{\includegraphics[width=0.46\columnwidth]{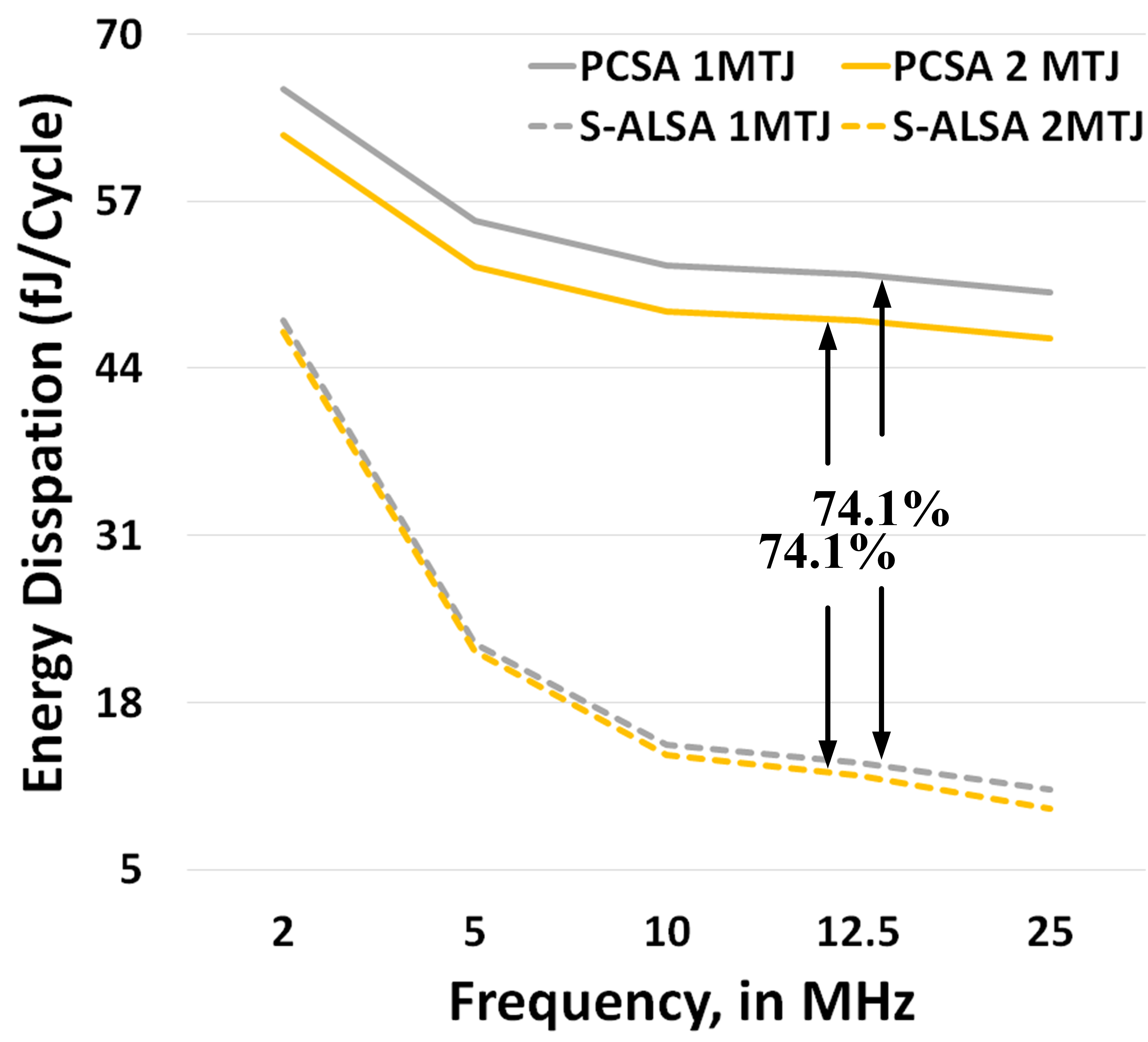}}
\caption{Energy dissipation vs.\ frequency for $4$ $\times$ $4$ MRAM macros.}
\label{fig:4x4_energy_f}
\end{figure}

Table~\ref{table:4x4_energy_f} lists energy dissipation at 2--25\,MHz, covering the operating range of IoT medical and sensor devices~\cite{9674039}. S-ALSA achieves up to 80.3\% energy saving over PCSA, with significant gains above 5\,MHz (Fig.~\ref{fig:4x4_energy_f}). Importantly, the balanced 4T-2MTJ bit cell incurs no energy penalty over 1T-1MTJ when paired with S-ALSA — at 12.5\,MHz, S-ALSA with 4T-2MTJ consumes 12.2\,fJ versus 13.8\,fJ with 1T-1MTJ in STT-MRAM, confirming that structural current balancing and energy efficiency are fully compatible in this co-design.

\begin{table*}[t]
\caption{Energy dissipation (fJ/cycle) of $4$ $\times$ $4$ MRAM macros at various frequencies.}
\renewcommand{\arraystretch}{1}
\setlength{\tabcolsep}{6pt}
\centering
\resizebox{\textwidth}{!}{
\begin{tabular}{cc ccccc ccccc}
\toprule
& & \multicolumn{5}{c}{$4$ $\times$ $4$ STT-MRAM} & \multicolumn{5}{c}{$4$ $\times$ $4$ SOT-STT-MRAM} \\
\cmidrule(lr){3-7} \cmidrule(lr){8-12}
Bit Cell & Sense Amp. & 2 MHz & 5 MHz & 10 MHz & 12.5 MHz & 25 MHz & 2 MHz & 5 MHz & 10 MHz & 12.5 MHz & 25 MHz \\
\midrule
\multirow{3}{*}{1T-1MTJ}
  & PCSA \cite{zhao2009high,9388516} & 66.8 & 59.1 & 55.5 & 54.8 & 53.3 & 65.8 & 55.5 & 52.0 & 51.3 & 49.9 \\
  & Proposed S-ALSA                 & 53.9 & 24.7 & 15.5 & 13.8 & 10.7 & 47.7 & 22.6 & 14.7 & 13.3 & 11.2 \\
  & Saving (\%)                     & 19.3 & 58.2 & 72.0 & 74.9 & 80.0 & 27.5 & 59.2 & 71.7 & 74.1 & 77.5 \\
\midrule
\multirow{3}{*}{4T-2MTJ}
  & PCSA \cite{zhao2009high,9388516} & 61.9 & 52.9 & 49.5 & 48.8 & 47.4 & 62.1 & 51.9 & 48.4 & 47.7 & 46.3 \\
  & Proposed S-ALSA                 & 49.4 & 22.3 & 13.8 & 12.2 & 9.32 & 46.8 & 22.0 & 13.9 & 12.4 & 9.69 \\
  & Saving (\%)                     & 20.2 & 57.9 & 72.1 & 75.1 & 80.3 & 24.6 & 57.7 & 71.3 & 74.1 & 79.1 \\
\bottomrule
\end{tabular}
}
\label{table:4x4_energy_f}
\end{table*}

\subsection{Energy vs. Temperature and TMR}

$4$ $\times$ $4$ macros were simulated at 12.5\,MHz across $-25^\circ$C to $75^\circ$C and TMR values from 150\% to 250\%, with results combined in Table~\ref{table:4x4_energy_temp}. S-ALSA achieves >82\% savings at $-25^\circ$C, remaining above 45\% at $75^\circ$C where increased PMOS leakage reduces the gain. Savings remain consistently near 75\% across the full TMR range, confirming robustness to MTJ process variation.

\begin{table*}[t]
\caption{Energy (fJ/cycle) vs.\ temperature (12.5\,MHz, TMR\,=\,200\%) and vs.\ TMR (12.5\,MHz, $V_{dd}$\,=\,1.2\,V) for $4$ $\times$ $4$ STT-MRAM and SOT-STT-MRAM macros.}
\renewcommand{\arraystretch}{1}
\centering
\setlength{\tabcolsep}{4pt}
\begin{tabular}{cc ccccc | ccccc}
\toprule
& & \multicolumn{5}{c|}{Temperature Sweep} & \multicolumn{5}{c}{TMR Sweep} \\
\midrule
& & $-25^\circ$C & $0^\circ$C & $25^\circ$C & $50^\circ$C & $75^\circ$C & 150\% & 175\% & 200\% & 225\% & 250\% \\
\midrule
\multicolumn{12}{c}{$4$ $\times$ $4$ STT-MRAM}\\
\midrule
\multirow{3}{*}{1T-1MTJ} & PCSA \cite{zhao2009high,9388516} & 51.8 & 53.1 & 54.8 & 57.3 & 61.0 & 56.4 & 55.5 & 54.8 & 54.3 & 53.8 \\
& S-ALSA                                                      & 9.19 & 10.3 & 13.8 & 20.2 & 31.3 & 14.31 & 14.01 & 13.8 & 13.6 & 13.5 \\
& Saving (\%)                                                 & 82.3 & 80.6 & 74.9 & 64.7 & 48.7 & 74.6 & 74.8 & 74.9 & 75.0 & 74.9 \\
\midrule
\multirow{3}{*}{4T-2MTJ} & PCSA \cite{zhao2009high,9388516} & 46.9 & 47.7 & 48.8 & 50.6 & 53.7 & 49.8 & 49.2 & 48.8 & 48.4 & 48.1 \\
& S-ALSA                                                      & 7.56 & 8.86 & 12.2 & 18.35 & 29.1 & 12.42 & 12.31 & 12.17 & 12.0 & 11.98 \\
& Saving (\%)                                                 & 83.9 & 81.4 & 75.1 & 63.8 & 45.9 & 75.1 & 75.0 & 75.1 & 75.1 & 75.1 \\
\midrule
\multicolumn{12}{c}{$4$ $\times$ $4$ SOT-STT-MRAM}\\
\midrule
\multirow{3}{*}{1T-1MTJ} & PCSA \cite{zhao2009high,9388516} & 49.6 & 50.5 & 51.8 & 53.8 & 57.0 & 52.2 & 51.7 & 51.3 & 51.0 & 50.7 \\
& S-ALSA                                                      & 8.98 & 9.90 & 13.0 & 19.12 & 29.4 & 13.50 & 13.39 & 13.30 & 13.2 & 13.14 \\
& Saving (\%)                                                 & 81.9 & 80.4 & 74.9 & 64.5 & 48.5 & 74.1 & 74.1 & 74.1 & 74.1 & 74.1 \\
\midrule
\multirow{3}{*}{4T-2MTJ} & PCSA \cite{zhao2009high,9388516} & 46.3 & 47.0 & 48.7 & 49.7 & 52.6 & 48.6 & 48.1 & 47.7 & 47.4 & 47.1 \\
& S-ALSA                                                      & 7.72 & 9.32 & 12.3 & 18.22 & 28.2 & 12.60 & 12.48 & 12.37 & 12.3 & 12.20 \\
& Saving (\%)                                                 & 83.3 & 80.2 & 74.3 & 63.4 & 46.3 & 74.0 & 74.1 & 74.1 & 74.1 & 74.1 \\
\bottomrule
\end{tabular}
\label{table:4x4_energy_temp}
\label{table:4x4_energy_TMR}
\end{table*}

\section{Security Analysis}\label{sec:security}

The 1T-1MTJ uses a fixed reference cell at the midpoint of $R_P$ and $R_{AP}$, so reading logic `0' draws more current than logic `1', producing a data-dependent current signature. This vulnerability is inherent to the cell structure and cannot be removed by the sense amplifier alone. The 4T-2MTJ stores data and its complement in two complementary MTJs; the sense amplifier always compares one P-state against one AP-state MTJ, structurally eliminating the asymmetry. When S-ALSA is paired with 4T-2MTJ, two mechanisms act together: the balanced cell removes storage-level leakage, and adiabatic logic suppresses residual dynamic power variation in the sensing circuit.

\begin{figure*}[t]
\centering
\subfigure[STT-MRAM, S-ALSA 1T-1MTJ (asymmetric)]{\includegraphics[width=0.15\textwidth]{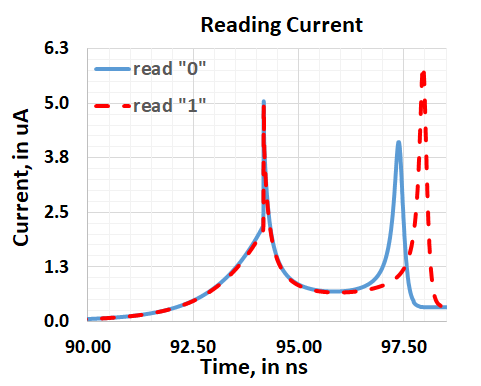}}
\subfigure[STT-MRAM, S-ALSA 4T-2MTJ (balanced)]{\includegraphics[width=0.15\textwidth]{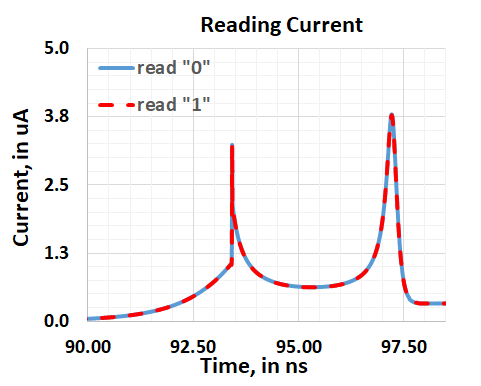}}
\hspace{4mm}
\subfigure[SOT-STT-MRAM, S-ALSA 1T-1MTJ (asymmetric)]{\includegraphics[width=0.15\textwidth]{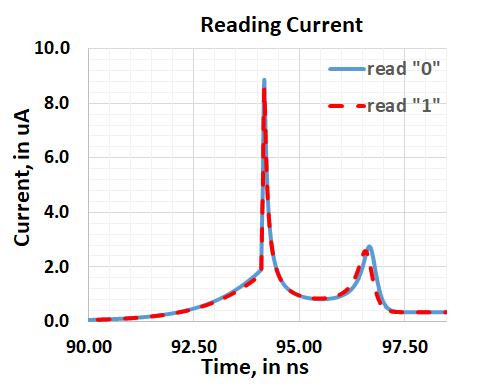}}
\subfigure[SOT-STT-MRAM, S-ALSA 4T-2MTJ (balanced)]{\includegraphics[width=0.15\textwidth]{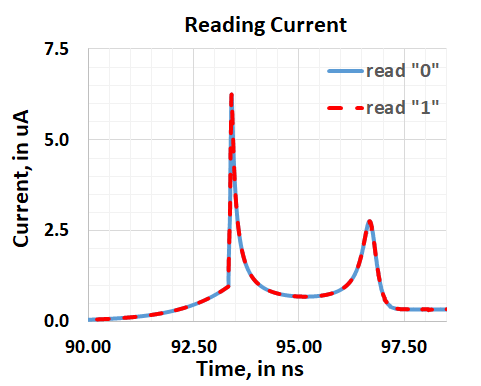}}
\caption{Read current of the S-ALSA 1-bit unit: STT-MRAM (a,b) and SOT-STT-MRAM (c,d). 1T-1MTJ shows asymmetric profiles; 4T-2MTJ shows balanced profiles across logic values.}
\label{fig:1bit_current}
\end{figure*}

Fig.~\ref{fig:1bit_current} confirms this at the 1-bit level. With 1T-1MTJ the read current differs visibly between logic values; with 4T-2MTJ the traces are nearly identical. The SOT-STT-MRAM 4T-2MTJ traces are even more tightly matched due to the decoupled SOT write path. The following subsections scale this analysis to $4$ $\times$ $4$ macros.

The following subsections scale this analysis to $4$ $\times$ $4$ macros using current traces, NED/NSD metrics, and CPA attacks.

\subsection{Current Trace Analysis}

\begin{figure*}[t]
\centering
\subfigure[STT, CMOS 1T-1MTJ]{\includegraphics[width=0.15\textwidth]{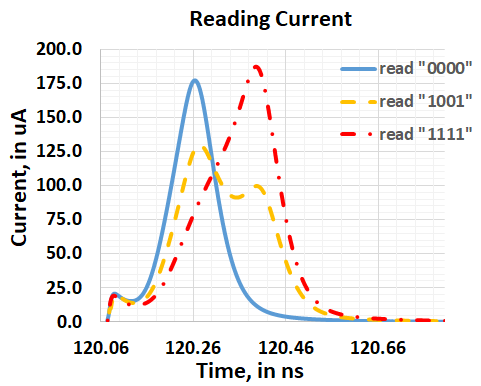}}
\subfigure[STT, S-ALSA 1T-1MTJ]{\includegraphics[width=0.15\textwidth]{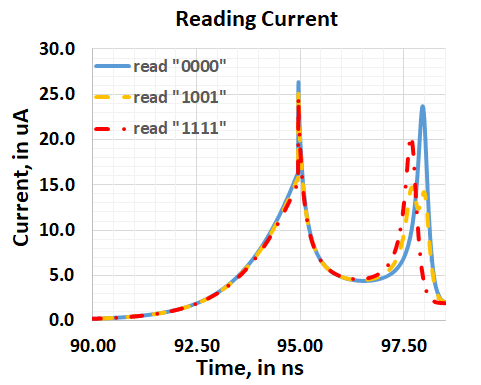}}
\subfigure[STT, CMOS 4T-2MTJ]{\includegraphics[width=0.15\textwidth]{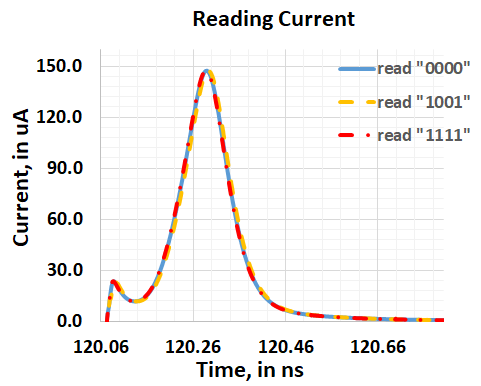}}
\subfigure[STT, S-ALSA 4T-2MTJ]{\includegraphics[width=0.15\textwidth]{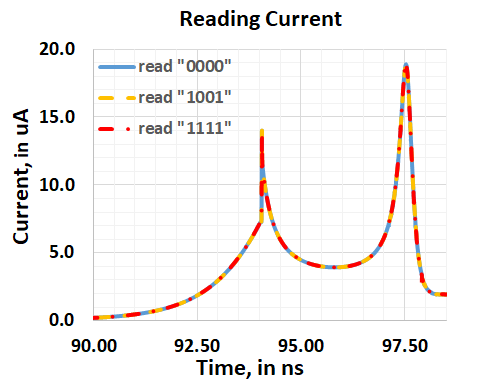}}
\\
\subfigure[SOT, CMOS 1T-1MTJ]{\includegraphics[width=0.15\textwidth]{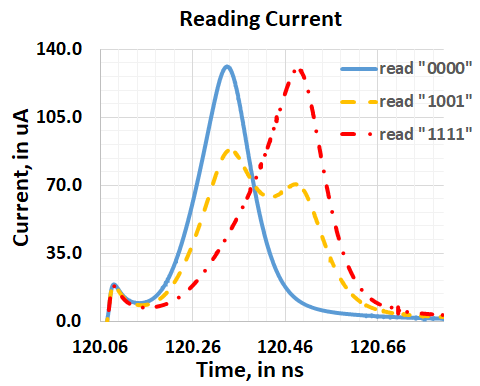}}
\subfigure[SOT, S-ALSA 1T-1MTJ]{\includegraphics[width=0.15\textwidth]{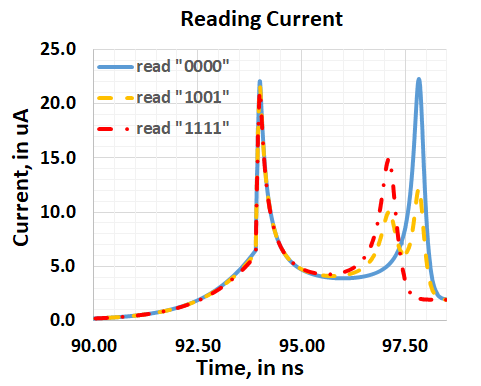}}
\subfigure[SOT, CMOS 4T-2MTJ]{\includegraphics[width=0.15\textwidth]{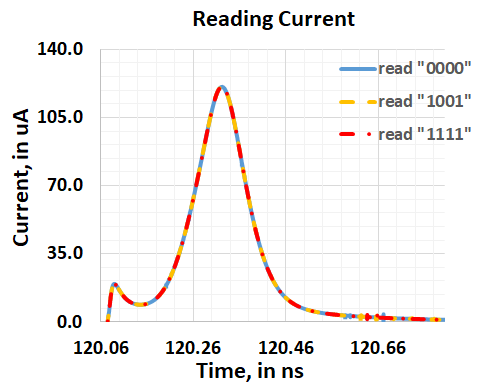}}
\subfigure[SOT, S-ALSA 4T-2MTJ]{\includegraphics[width=0.15\textwidth]{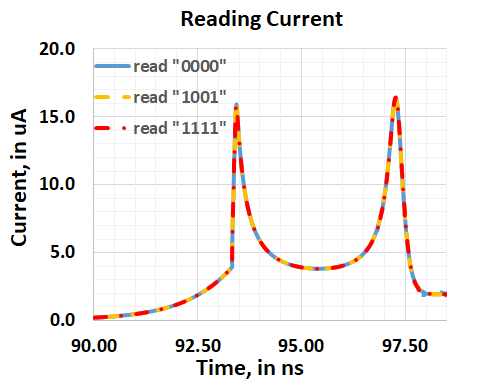}}
\caption{Read current traces of $4$ $\times$ $4$ STT-MRAM (a--d) and SOT-STT-MRAM (e--h). 1T-1MTJ (a,b,e,f) shows distinguishable profiles; 4T-2MTJ (c,d,g,h) shows balanced profiles confirming security at the circuit level.}
\label{fig:CSTT_1T}
\label{fig:CSTT_2T}
\end{figure*}

We collected read current traces for all 16 stored-value combinations in the $4$ $\times$ $4$ MRAM macros for both STT-MRAM and SOT-STT-MRAM. For 1T-1MTJ (Fig.~\ref{fig:CSTT_1T}a,b,e,f), both PCSA and S-ALSA exhibit distinguishable current profiles between logic `0' and `1' across both technologies, consistent with the inherent bit cell asymmetry. For 4T-2MTJ (Fig.~\ref{fig:CSTT_1T}c,d,g,h), both designs show nearly identical traces for all 16 stored values in both STT-MRAM and SOT-STT-MRAM, confirming that the structural balance of the complementary MTJ pair eliminates the data-dependent signature at the circuit level. Notably, the SOT-STT-MRAM 4T-2MTJ traces are even more tightly matched than their STT-MRAM counterparts, consistent with the decoupled SOT write path reducing residual read-current asymmetry between the two MTJs.

\subsection{NED and NSD Evaluation}

To quantify leakage, we compute Normalized Energy Deviation (NED) and Normalized Standard Deviation (NSD) across all 16 stored-value combinations~\cite{NEDNSD}. Lower values indicate less leakage and greater CPA resistance.

\begin{equation}
NED = \frac{E_{max} - E_{min}}{E_{max}}, \quad NSD = \frac{\sigma_E}{E_{avg}}
\label{equ:NED_NSD}
\end{equation}

STT-MRAM (Table~\ref{table:NED_STT}): With PCSA, switching from 1T-1MTJ to 4T-2MTJ alone reduces NED and NSD by 67.7$\times$ and 95.4$\times$ respectively, confirming that bit cell symmetry is the dominant balancing factor. S-ALSA with 1T-1MTJ shows NED/NSD comparable to PCSA with 1T-1MTJ (11.67\% vs.\ 11.58\%), as expected given the bit cell asymmetry. With 4T-2MTJ, S-ALSA achieves a further 5.44$\times$ and 5.64$\times$ reduction in NED and NSD over PCSA with the same bit cell, confirming that the sense amplifier contributes an additional, independent layer of protection on top of the balanced bit cell.

SOT-STT-MRAM (Table~\ref{table:NED_SOT}): Comparing S-ALSA with 4T-2MTJ against S-ALSA with 1T-1MTJ in the SOT-STT-MRAM configuration reveals a 2278$\times$ reduction in NED (from 6.15\% down to 0.0027\%) and a 2271$\times$ reduction in NSD (from 1.59\% down to 0.0007\%). This result demonstrates that within the SOT-STT-MRAM technology, the choice of bit cell has an overwhelming impact on security: both designs use the same adiabatic sense amplifier, yet switching from 1T-1MTJ to 4T-2MTJ collapses the leakage to near-zero. This stems from a unique SOT-STT-MTJ property: the SOT write path is decoupled from the read path via a separate heavy-metal layer, which reduces residual read-current asymmetry between the two MTJs in the 4T-2MTJ cell. Consequently, the 4T-2MTJ structure becomes even more symmetric under SOT operation, reducing energy variance across all 16 stored-value combinations toward near-zero. This makes SOT-STT-MRAM with 4T-2MTJ the preferred configuration when security is the primary design concern.

\begin{table*}[t]
\caption{NED and NSD values for $4$ $\times$ $4$ STT-MRAM (left) and SOT-STT-MRAM (right) at 12.5\,MHz.}
\renewcommand{\arraystretch}{1}
\setlength{\tabcolsep}{5pt}
\centering
\begin{tabular}{c cccc | cccc}
\toprule
& \multicolumn{4}{c|}{$4$ $\times$ $4$ STT-MRAM} & \multicolumn{4}{c}{$4$ $\times$ $4$ SOT-STT-MRAM} \\
\midrule
& \multicolumn{2}{c}{PCSA \cite{zhao2009high,9388516}} & \multicolumn{2}{c|}{Proposed S-ALSA}
& \multicolumn{2}{c}{PCSA \cite{zhao2009high,9388516}} & \multicolumn{2}{c}{Proposed S-ALSA} \\
\cmidrule(lr){2-3}\cmidrule(lr){4-5}\cmidrule(lr){6-7}\cmidrule(lr){8-9}
Parameter & 1T-1MTJ & 4T-2MTJ & 1T-1MTJ & 4T-2MTJ & 1T-1MTJ & 4T-2MTJ & 1T-1MTJ & 4T-2MTJ \\
\midrule
$E_{max}$ (fJ) & 112.6 & 95.8 & 22.4 & 19.3 & 106.2 & 94.0 & 27.7 & 25.6 \\
$E_{min}$ (fJ) & 99.6  & 95.7 & 19.8 & 18.9 & 97.2  & 93.9 & 26.0 & 25.6 \\
$E_{avg}$ (fJ) & 106.2 & 95.7 & 21.3 & 19.2 & 101.7 & 93.9 & 26.9 & 25.6 \\
$\sigma_E$      & 3.28E-15 & 3.10E-17 & 7.04E-16 & 1.12E-16 & 2.24E-15 & 1.55E-17 & 4.26E-16 & 1.80E-19 \\
NED (\%)        & 11.58 & 0.17  & 11.67 & 2.14   & 8.48  & 0.07  & 6.15  & 0.0027 \\
NSD (\%)        & 3.08  & 0.03  & 3.30  & 0.59   & 2.21  & 0.02  & 1.59  & 0.0007 \\
\bottomrule
\end{tabular}
\label{table:NED_STT}
\label{table:NED_SOT}
\end{table*}

\subsection{CPA Attack Validation}

CPA attacks were performed on $4$ $\times$ $4$ MRAM macros assuming they store PRESENT-80 intermediate cipher values during encryption rounds~\cite{STTsca, wu2012measurement}. For each plaintext, the attacker measures the supply current during the MRAM read cycle, builds a Hamming-weight hypothesis for each candidate key, and computes the Pearson correlation between measured and hypothetical power traces. A divide-and-conquer approach extracts each 4-bit subkey independently. A correlation above 0.9 for the correct key with all others below 0.5 declares a successful attack.

\begin{figure*}[t]
\centering
\subfigure[STT-MRAM, CMOS 1T-1MTJ (succeeds)]{\includegraphics[width=0.19\textwidth]{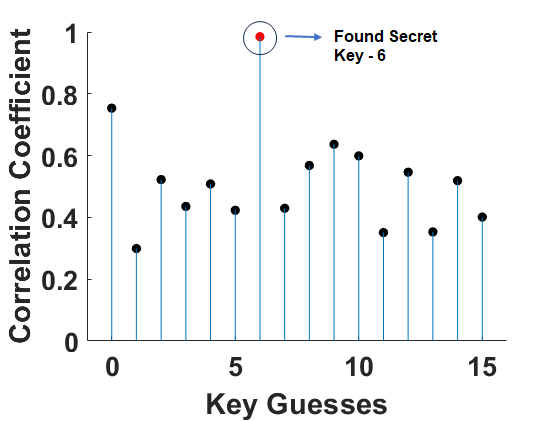}}
\subfigure[STT-MRAM, S-ALSA 4T-2MTJ (fails)]{\includegraphics[width=0.19\textwidth]{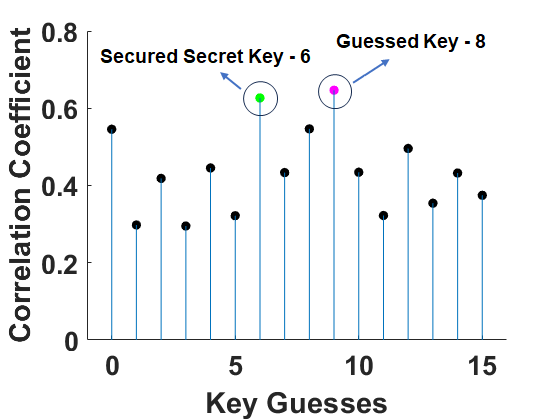}}
\hspace{4mm}
\subfigure[SOT-STT-MRAM, CMOS 1T-1MTJ (succeeds)]{\includegraphics[width=0.19\textwidth]{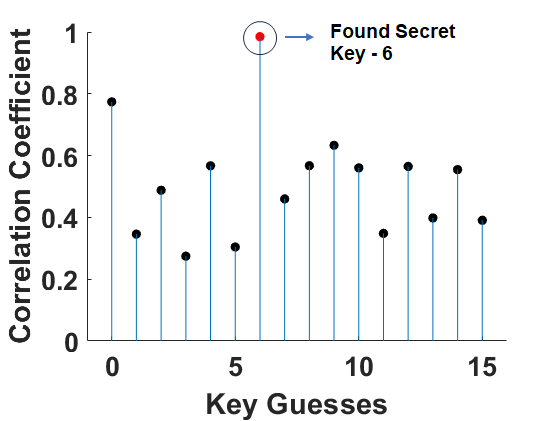}}
\subfigure[SOT-STT-MRAM, S-ALSA 4T-2MTJ (fails)]{\includegraphics[width=0.19\textwidth]{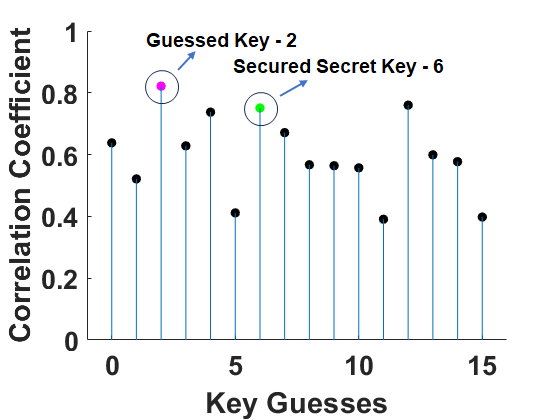}}
\caption{CPA attack on $4$ $\times$ $4$ STT-MRAM (a,b) and SOT-STT-MRAM (c,d) for PRESENT-80 encryption. Attack succeeds on 1T-1MTJ (a,c); attack fails on 4T-2MTJ (b,d).}
\label{fig:CPA_STT}
\label{fig:CPA_SOT}
\end{figure*}

Fig.~\ref{fig:CPA_STT} shows CPA results for both STT-MRAM (a,b) and SOT-STT-MRAM (c,d). With 1T-1MTJ, both PCSA and S-ALSA are successfully attacked: correct key 6 (red dot) achieves peak correlation $\approx$0.985. S-ALSA with 1T-1MTJ is also vulnerable, confirming the adiabatic sense amplifier alone cannot compensate for fundamental bit cell asymmetry. With 4T-2MTJ, the attack fails for both technologies: the correct key (green dot) is indistinguishable from wrong guesses (purple dots), with all correlations near 0.1. The bit cell provides structural current balance while adiabatic logic suppresses residual dynamic power variation. Together, S-ALSA and the 4T-2MTJ bit cell function as co-designed security partners, delivering a robust, energy-efficient, and attack-resistant memory reading solution.

\section{Conclusion}\label{sec:conclusion}

This paper presents S-ALSA, a Secured Adiabatic Logic Sense Amplifier that extends prior energy-efficient MRAM design to include side-channel resistance. The central contribution is a co-design framework combining two independent protection mechanisms:structural current balancing through the 4T-2MTJ bit cell and dynamic power equalization through adiabatic charge recovery in the sensing circuit. The 4T-2MTJ cell incurs approximately $3.5\times$ area overhead relative to 1T-1MTJ; however, this tradeoff is favorable for security-critical IoT applications where side-channel resistance is a design requirement. The proposed design achieves up to 80\% energy reduction over conventional PCSA designs, effectively suppresses observable data-dependent power leakage in both STT-MRAM and SOT-STT-MRAM, and defeats CPA key extraction as confirmed by PRESENT-80 attack experiments, without any energy penalty. These results hold across IoT-relevant operating conditions of temperature and MTJ process variation. The current security evaluation focuses on passive power-analysis attacks using a sense resistor, the most practical non-invasive attack model for IoT deployments. This work establishes a new paradigm where energy efficiency and hardware security are jointly optimized through coordinated design across device and circuit levels, making S-ALSA highly suitable for next-generation IoT memory systems.

\balance
\section*{Acknowledgment}
This work is partially supported by National Science Foundation CAREER Award No. 2607625.

{\small
\setlength{\itemsep}{0pt}
\setlength{\parsep}{0pt}
\bibliographystyle{IEEEtran}
\bibliography{Reference}
}

\end{document}